# TRANSPORT CURRENTS AND PERSISITENT CURRENTS IN SOLIDS: ORBITAL MAGNETISM AND HALL EFFECT OF DIRAC ELECTRONS

Hidetoshi Fukuyama[1], Yuki Fuseya[2] and Akito Kobayashi[3]

[1]*Department of Applied Physics and Research Institute for Science and Technology Tokyo University of Science
Kagurazaka, Shinjuku-ku, Tokyo 162-860, Japan
E-mail:fukuyama@rs.kagu.tus.ac.jp*

[2]*Division of Materials Physics, Department of Materials Engineering Science, Graduate School of Engineering Science, Osaka University
Toyonaka, Osaka 560-8531*

[3]*Institute for Advanced Research, Nagoya University
Furo-cho, Chikusa-ku, Nagoya 464-8602*

Features of electronic currents in solids are truly diverse depending on circumstances, e.g. non-equilibrium transport currents leading to dissipation and persistent currents flowing in equilibrium. Differences between these currents may be clear in many cases, while there are some where they are not. Results of theoretical studies on the latter cases will be introduced briefly focusing on the inter-band effects of magnetic fields in orbital magnetisms and Hall effects of Dirac electrons.

1. Introduction

Orbital magnetism is due to currents flowing in equilibrium state, i.e. persistent currents, caused by external magnetic field. It is clear that such currents flow even in insulators as is evidenced by the existence of atomic diamagnetism. On the other hand, the celebrated Landau diamagnetism is for conduction electrons. The relationship between these in solids, where energy spectra of electrons are described by Bloch bands leading to metallic or insulating states depending on the location of the Fermi energy, has not been explored in full details. At the same time, the Hall effect in weak magnetic field, which reflects the changes of electronic transport current caused by the field, is non-equilibrium properties in the presence of finite electric field. The relationship between these two phenomena, orbital magnetism and Hall effect, has not been explored so far. In this short note, some of recent results of studies on this problem are reported.

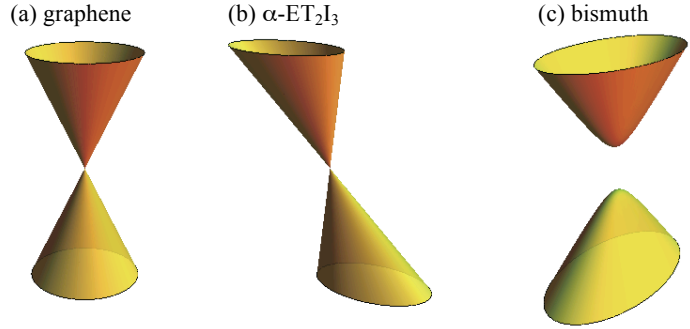

Fig. 1. Energy spectra of electrons for (a) graphene, (b) α-ET$_2$I$_3$ and (c) bismuth.

## 2. Inter-band Effects of Magnetic Field on Dirac Electrons

The simplest and most transparent way to take account of magnetic field for electrons in solids, i.e. Bloch electrons, is to introduce the Peierls phase in the transfer integral. The orbital susceptibility for Bloch electrons based on this approximation is given by the Landau-Peierls (LP) formula[1]. However in this approximation the fact that the vector potential representing magnetic field has finite matrix elements between Bloch bands in any circumstances is totally ignored. Actually LP formula does not reflect the effects of weak periodic potential even in its second order[2]. More dramatically theoretical results based on LP formula are in complete disagreement with the experimental observation of large diamagnetism in semimetals, such as graphite and bismuth. It has been clarified that the inter-band effects of magnetic field play crucial roles for the understanding of large diamagnetism in both of these semimetals[3,4]. Natural question then is on the possible consequences of such inter-band effects of magnetic field on the Hall effect, which has been studied recently[5-8]. Actual systems of interest are two-dimensional graphenes and molecular solids, $\alpha$ET$_2$I$_3$, and bulk crystal of bismuth. It is to be noted that these are described as Dirac electrons[9-11]. The former two, graphenes and $\alpha$ET$_2$I$_3$, are described by 2x2 Weyl equation for massless Dirac electrons (Dirac cones, whose energy and wave number at the tips may be called "crossing energy" and "crossing point") similarly but with distinct differences between them: there is a finite tilting of the cones in α-ET$_2$I$_3$ since their crossing points are located at off-symmetry points in the Brillouin zone[12]. On the other hand the proper model for bismuth, where spin-orbit interactions are very strong, is 4x4 Dirac equations with spatial anisotropy of velocity. The energy spectra of electrons described by these Dirac equations are shown in Fig, 1(a) for graphenes, in Fig 1(b) for α-ET$_2$I$_3$ and in Fig 1(c) for bismuth, respectively.

In order to study such subtle inter-band effects of magnetic field on orbital susceptibility and Hall effect on firm ground, the exact formulas for orbital susceptibility, $\chi$, and Hall conductivity, $\sigma_{xy}$, are employed which are derived by use of the Luttinger-Kohn representation[13] suited to identify the gauge-invariance associated with vector potential[14,15]. They are given as follows.

$$\chi = \frac{e^2}{c^2} T \sum_n \sum_k \text{Tr} G\gamma_x G\gamma_y G\gamma_x G\gamma_y, \quad (1)$$

$$\sigma_{\mu\nu} = \frac{1}{i\omega} K^\alpha_{\mu\nu}(q,\omega) A_{q\alpha}, \quad (2)$$

$$K^\alpha_{\mu\nu} = -\frac{e^3}{2mc}\left(q_\mu \delta_{\nu\alpha} - q_\nu \delta_{\mu\alpha}\right) T \sum_n \sum_k \text{Tr}\left[G_-\gamma_\mu G\gamma_\mu G - G_-\gamma_\mu G_-\gamma_\mu G\right]$$

$$-\frac{e^3}{2c}\left(q_\mu \delta_{\nu\alpha} - q_\nu \delta_{\mu\alpha}\right) T \sum_n \sum_k \text{Tr}\left[G_-\gamma_\nu G_-\gamma_\mu G\gamma_\mu G\gamma_\nu - G_-\gamma_\mu G_-\gamma_\mu G\gamma_\nu G\gamma_\nu \right. \quad (3)$$

$$+\gamma_\mu G\gamma_\nu G_-\gamma_\mu G_-\gamma_\nu G_- - \gamma_\mu G\gamma_\nu G_-\gamma_\nu G_-\gamma_\mu G_-$$

$$\left. +G_-\gamma_\mu G\gamma_\mu G\gamma_\nu G\gamma_\nu - G_-\gamma_\mu G\gamma_\nu G\gamma_\mu G\gamma_\nu \right]$$

where $\gamma$ is the velocity matrix $\gamma_\mu = \partial H/\partial k_\mu$ the Green function is given by $G = [i\tilde{\varepsilon}_n - H + \mu]^{-1}$ ( $G_- = [i\tilde{\varepsilon}_n - i\omega_m - H + \mu]^{-1}$ ) with $\tilde{\varepsilon}_n = \varepsilon_n + \Gamma\text{sgn}(\varepsilon_n)$ ( $\varepsilon_n = (2n+1)\pi T$ ), μ and Γ being the chemical potential and the spectrum broadening, respectively, and the vector potential with finite Fourier component $q$, $A_q$, is introduced to represent even spatially uniform magnetic field $B$ to make gauge invariance visible by taking the limit of $q \to 0$, i.e. $i\boldsymbol{q} \times \boldsymbol{A}_q = \boldsymbol{B}$.

## 3. Graphenes and $\alpha$-ET$_2$I$_3$

The general model applicable to both graphenes and molecular solids $\alpha$-ET$_2$I$_3$ is given as follows[10]

$$H = \sum_{\mu=0,1,2,3} \boldsymbol{k} \cdot \boldsymbol{V}_\mu \sigma_\mu. \qquad (4)$$

where $\sigma_{1,2,3}$ are the Pauli matrices and $\sigma_0$ is the identity matrix. The momentum $\boldsymbol{k}$ is measured from the crossing point. This has been deduced from the effective Hamiltonian describing the motion of electrons around the crossing point, and then it is rigorous in the vicinity of the crossing point. The band energy dispersion is given as follows

$$E = \boldsymbol{k} \cdot \boldsymbol{V}_0 \pm \sqrt{\sum_{\mu=x,y,z} (\boldsymbol{k} \cdot \boldsymbol{V}_\mu)^2}. \qquad (5)$$

In graphenes, the vector $\boldsymbol{V}_0 = \boldsymbol{0}$ and the velocity of cones is isotropic, *i. e.* $\boldsymbol{V}_x = (v, 0)$, $\boldsymbol{V}_y = (0, v)$, and $\boldsymbol{V}_z = \boldsymbol{0}$. In $\alpha$-ET$_2$I$_3$. the velocity strongly depends on the direction of the motion. The vectors $\boldsymbol{V}_x$, $\boldsymbol{V}_y$, $\boldsymbol{V}_z$ result in the anisotropy of cones. The vector $\boldsymbol{V}_0$ tilts the axis of the cones, which makes difference between the velocities in a direction and in the opposite direction. Based on the band calculation, we can take $\boldsymbol{V}_0 = (v_0, 0)$, $\boldsymbol{V}_x = (v_x, 0)$, $\boldsymbol{V}_y = 0$, $\boldsymbol{V}_z = (0, v_z)$ with $v_0 = 0.8 \times 10^5$ m/s and $v_x = v_z = 1.0 \times 10^5$ m/s for analytical calculations in the following. In this case the highest velocity is about 10 times larger than the lowest one in the opposite direction.

The results of calculations of $\sigma_{xy}$ and $\chi$ based on eqs. (1)-(4) are shown in Figs. 2 (a) and (b) for $\alpha$-ET$_2$I$_3$ together with those for graphenes as a special case as a function of $X = \mu / \Gamma$.

The orbital diamagnetism in Fig. 2 (b) has large values around the crossing energy and strong $\mu$-dependences in the region of $|X| < 1$. The orbital diamagnetism in $\alpha$-ET$_2$I$_3$ exhibits the same $X$-dependence as that of graphene, but is enhanced by tilting.

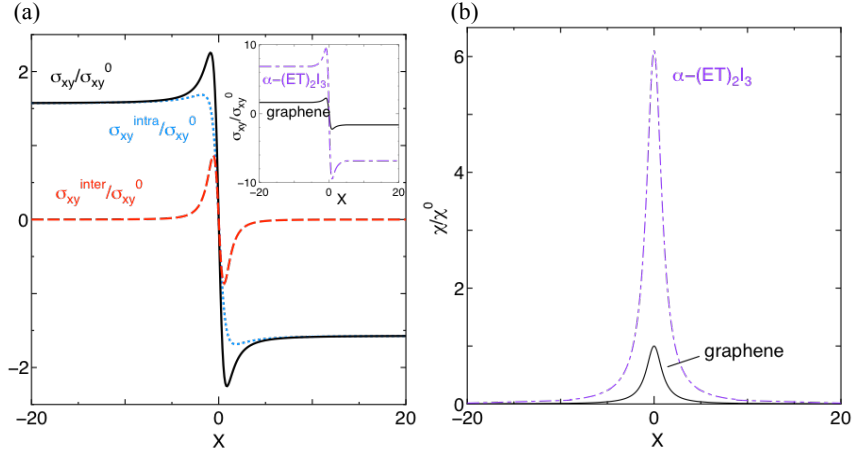

Fig.2. Dependences on chemical potential, $\mu$, (scaled by the damping energy $\Gamma$, i.e. X= $\mu/\Gamma$) of Hall conductivity (a) and orbital susceptibility (b) in the absence of tilting (graphenes) and for α-ET$_2$I$_3$ with the choices of $v_0$=0.8x10$^5$m/s and v=1.0x10$^5$m/s deduced from band calculations.

In Fig. 2(a), the Hall conductivity σ$_{xy}$ (the solid line) is given by the sum of σ$_{xy}^{inter}$ (the dashed red line) and σ$_{xy}^{intra}$ (the dotted blue line), which are defined as contributions from all the states below the Fermi energy and those at the Fermi energy, respectively. These two terms increase with increasing tilting, although they exhibit different tilting-dependences with each other. It is seen that the contributions to σ$_{xy}^{inter}$ are confined in energy region where the orbital susceptibility takes large values.

An important fact Fig. 2 (a) indicates is that the Hall coefficient, $R_H$ is vanishing and changing signs at the crossing energy. This implies that the convention to deduce the effective carrier density, *n*, by $n = (ecR_H)^{-1}$ is totally invalid, since *n* = 0 is expected if the Fermi energy is located at the crossing energy. Another interesting feature expected for α-ET$_2$I$_3$, whose energy spectra relative to the crossing energy is not symmetric, i.e. without "electron-hole symmetry", is that the chemical potential is expected to depend on temperature and can pass though the crossing energy at low temperature in the presence of finite but very small amount (even of the order of ppm) of doped carriers. If this happens, the Hall coefficient can undergo very sharp change of sign as a function of chemical potential as shown in Fig.3 (c). Such is actually been observed experimentally[15].

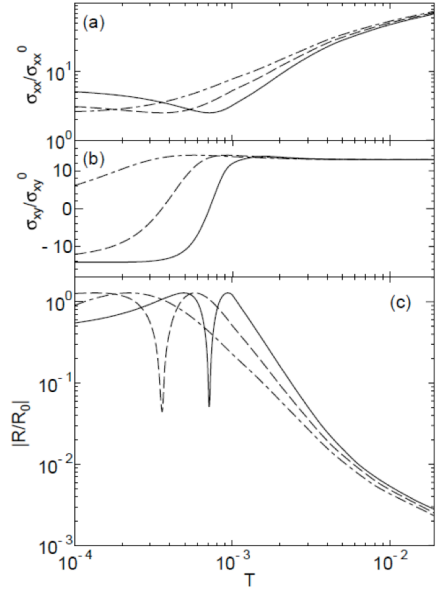

Fig. 3. Temperature dependence of (a) $\sigma_{xx}$, (b) $\sigma_{xy}$, and (c) the Hall coefficient, where the solid, dashed, and dot-dashed lines are calculated for finite doping expressed as the chemical potential off the crossing energy at absolute zero $\Delta\mu = 2.0 \times 10^{-4}$, $1.0 \times 10^{-4}$, and 0, respectively.

## 4. Bismuth

The effective model for Bi is given by[11]

$$H = \frac{E_G}{2}\beta + i\bm{k}\cdot\left[\sum_{\mu=1}^{3} W(\mu)\beta\alpha_\mu\right]$$

$$= \begin{pmatrix} E_G/2 & 0 & \bm{k}\cdot\langle 1|v|3\rangle & \bm{k}\cdot\langle 1|v|4\rangle \\ 0 & E_G/2 & \bm{k}\cdot\langle 2|v|3\rangle & \bm{k}\cdot\langle 2|v|4\rangle \\ \bm{k}\cdot\langle 3|v|1\rangle & \bm{k}\cdot\langle 3|v|2\rangle & -E_G/2 & 0 \\ \bm{k}\cdot\langle 4|v|1\rangle & \bm{k}\cdot\langle 4|v|2\rangle & 0 & -E_G/2 \end{pmatrix}, \qquad (6)$$

where $E_G$ is the band gap, and $\beta$, $\alpha_\mu$ are the 4x4 matrices that appear in the Dirac theory. The quantities $\langle i|v|j\rangle$ are the matrix elements of the velocity operator, and the indexes $i,j$ denote the four band-edge wave function (2 for bands and 2 for spins). We have measured energies from the center of the band gap. It is to be noted that this model naturally leads to small effective mass and corresponding large g-factor of spin Zeeman splitting of states near the band gaps[11,17]. In order to see the essence of inter-band contributions this model is simplified by assuming the isotropic velocity $v$ as follows:

$$H = \frac{E_G}{2} + iv \sum_\mu k_\mu \beta \alpha_\mu. \tag{7}$$

The Hall conductivity, $\sigma_{xy}$, and the orbital susceptibility, $\chi$, are calculated on the basis of the exact formulas as in the previous section. The final expressions are

$$\sigma_{xx} = -\frac{e^2}{\pi^3 v} \int_{-\infty}^{\infty} d\varepsilon \int_0^{\infty} dX \left[ F_1(\varepsilon,X) - F_2(\varepsilon,X) \right] f'(\varepsilon - \mu), \tag{8}$$

$$F_1(\varepsilon,X) = \frac{X^2 \left( \varepsilon^2 + \Gamma^2 - X^2/3 - E_G^2/4 \right)}{\left\{ (\varepsilon + i\Gamma)^2 - X^2 - E_G^2/4 \right\} \left\{ (\varepsilon - i\Gamma)^2 - X^2 - E_G^2/4 \right\}},$$

$$F_2(\varepsilon,X) = \frac{X^2 \left\{ (\varepsilon + i\Gamma)^2 - X^2/3 - E_G^2/4 \right\}}{2 \left\{ (\varepsilon + i\Gamma)^2 - X^2 - E_G^2/4 \right\}^2} + \text{c.c.}, \tag{9}$$

$$\sigma_{xy} = \frac{e^3 vB}{12\pi^2 c} \int_{-\infty}^{\infty} d\varepsilon \left[ F_3(\varepsilon) f(\varepsilon - \mu) + F_4(\varepsilon) f'(\varepsilon - \mu) \right] \text{sgn}(\varepsilon), \tag{10}$$

$$F_3(\varepsilon) = \frac{\varepsilon + i\Gamma}{\left[ (\varepsilon + i\Gamma)^2 - E_G^2/4 \right]^{3/2}} + \text{c.c.},$$

$$F_4(\varepsilon) = \frac{\left( E_G^2/4 - \varepsilon^2 \right)^2 - 2\Gamma^4 - \Gamma^2 E_G^2/4 + 2i\Gamma^3 \varepsilon - i\Gamma\mu \left( E_G^2/4 - \varepsilon^2 \right)}{2\Gamma^2 \varepsilon^2 \sqrt{\varepsilon^2 - \Gamma^2 - E_G^2/4 + 2i\Gamma\varepsilon}} + \text{c.c.}, \tag{11}$$

$$\chi = \frac{4}{15} \frac{e^2 v}{c^2 \pi^2} \int_{-\infty}^{\infty} d\varepsilon \left[ \frac{1}{\sqrt{(\varepsilon + i\Gamma)^2 - E_G^2/4}} + \text{c.c.} \right] f(\varepsilon - \mu) \text{sgn}(\varepsilon). \tag{12}$$

The Hall coefficient is defined as $R_H = \sigma_{xy}/B\sigma_{xx}^2$.

The results of calculations of $\sigma_{xx}$, $\sigma_{xy}$, $R_H$, $\sigma_{xy}^{\text{inter}}$ and $\chi$ for this model as a function of the chemical potential, $\mu$, are shown in Fig. 4(a)-(e). Here, $\sigma_{xx0} = e^2/\pi^2 v$, $\sigma_{xy0} = e^3 v/12\pi^2 c$, $\chi_0 = 4e^2 v/15c^2\pi^2$.

The inter-band contributions are obtained by subtracting the intra-band contribution, $\sigma_{xy}^{\text{intra}}$, from the exact value $\sigma_{xy}$: $\sigma_{xy}^{\text{inter}} = \sigma_{xy} - \sigma_{xy}^{\text{intra}}$. The intra-band contribution can be calculated within the Bloch band picture as[14]

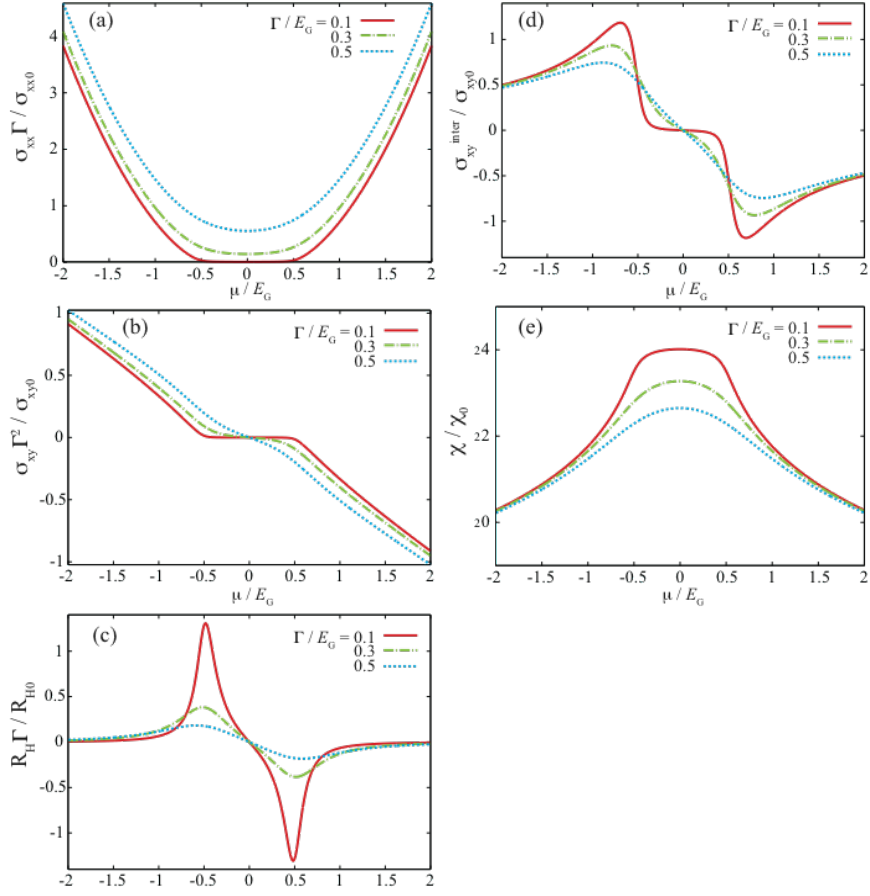

Fig. 4. The results of calculations for (a) the conductivity σ$_{xx}$, (b) the Hall conductivity σ$_{xy}$, (c) the Hall coefficient, (d) inter-band contribution σ$_{xy}^{\text{inter}}$, (e) the orbital susceptibility χ, as a function of chemical potential μ.

$$\sigma_{xy}^{\text{intra}} = -\frac{4}{3}\frac{e^3 B}{\pi c}\sum_k \int_{-\infty}^{\infty} dx \left[(\partial_x \xi)^2 \partial_y^2 \xi - \partial_x \xi \partial_y \xi \partial_{xy}\xi\right] \quad (13)$$

where $\xi = \sqrt{v^2 k^2 + E_G^2/4}$ and $\partial_\mu \xi = \partial \xi / \partial k_\mu$. Then we have the final expression for the present model:

$$\sigma_{xy}^{\text{intra}} = -\frac{e^3 vB}{6\pi^3 c}\sum_{n=\pm}\int_{-\infty}^{\infty}d\varepsilon \int_0^{\infty}dX \frac{nX^4}{[E_n(X)]^3}\frac{4\Gamma^3}{3\left[\{\varepsilon - E_n(X)\}^2 + \Gamma^2\right]^3} f'(\varepsilon - \mu), \quad (14)$$

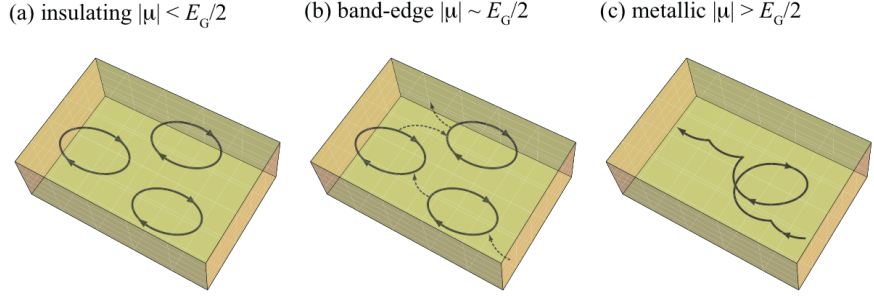

(a) insulating $|\mu| < E_G/2$  (b) band-edge $|\mu| \sim E_G/2$  (c) metallic $|\mu| > E_G/2$

Fig. 5. Schematic motion of electrons in a solid for (a) the insulating ($|\mu| < E_G/2$), (b) the band-edge ($|\mu| \sim E_G/2$), and (c) the metallic region ($|\mu| > E_G/2$).

where $E_\pm(X) = \pm\sqrt{X^2 + E_G^2/4}$.

It is seen that $\sigma_{xy}^{inter}$ exist near the band edges (Fig. 4 (d)), where $\chi$ (Fig. 4 (e)) takes large values except for insulating region. This strongly suggests that the nature of $\sigma_{xy}^{inter}$ is closely related to that of orbital current. The physical picture of this correlation can be understood as follows as schematically shown in Fig.5. In the insulating region, the current flows locally under a magnetic field, generating the diamagnetism (Fig. 5 (a)). This current is non-dissipative, so that it does not contribute to Hall currents. In the band-edge region, on the other hand, the local orbital currents hybridize with the conduction electrons, which will lead to the mixing between diamagnetic currents and Hall currents. This will be the origin of the $\sigma_{xy}^{inter}$ in the band-gap region.

In the clean limit, i.e. $\Gamma \to 0$, it is analytically seen that $\sigma_{xy}^{inter}$ is vanishingly small for $-0.5 < \mu/E_G < 0.5$ (band gap region). In this energy region orbital susceptibility shows maximum diamagnetism as seen in Fig. 4 (e). This implies that orbital currents leading to susceptibility are not contributing to Hall effect, and they are totally independent if the system is clean. Once disorder is introduced, e.g. by impurities, however, current flow will get intermixed and clear separation of these will be no longer clear because of finite $\sigma_{xy}$ in the band gap region. To identify characteristic features of current flows in such cases, careful studies are needed on the impurity states which have been seen to anomalously large g-factor[18] because of strong spin-orbit interactions. This coupling between orbital currents and spins is also an interesting subject in the context of spin-Hall effect[19], which deserves detailed studies.

**Acknowledgments**

One of authors (HF) is very thankful for Joe Imry for stimulating and enlightening discussions at various stages of researches on quantum transport properties in solids.